# Thomson scattering diagnostics at the Globus-M2 tokamak.

Zhiltsov N.S.[1], Kurskiev G.S.[1], Tolstyakov S.Yu.[1], Solovey V.A.[2], Koval A.N.[1], Aleksandrov S.E.[1], Bazhenov A.N.[1], Chernakov P.V.[3], Filippov S.V.[1], Gusev V.K.[1], Khromov N.A.[1], Kiselev E.O.[1], Kornev A.F.[4], Krikunov S.V.[1], Makarov A.M.[5], Minaev V.B.,[1] Miroshnikov I.V.[1], Mukhin E.E.[1], Novokhatsky A.N.[1], Patrov M.I.[1], Petrov Yu.V.[1], Sakharov N.V.[1], Schegolev.P.B.[1], Telnova A.Yu.[1], Tkachenko E.E.[1], Tokarev V.A.[1], Varfolomeev V.I.[1], Voronin A.V[1].

[1]*Ioffe Institute, Saint-Petersburg, Russia*

[2]*Petersburg Nuclear Physics Institute named by B.P. Konstantinov of NRC, "Kurchatov Institute", Saint-Petersburg, Russia*

[3]*Spectral-Tech JSC, Saint-Petersburg, Russia*

[4]*Lasers & Optical Systems Co. Ltd., Saint-Petersburg, Russia*

[5]*ITMO University., Saint-Petersburg, Russia*

E-mail: nisovru@gmail.com

*The paper is devoted to the **T**homson **s**cattering (TS) diagnostics recently developed for the Globus-M2 spherical tokamak and prototyping the ITER divertor TS diagnostics. The distinctive features of the system are the use of spectrometers, acquisition system and lasers that meet the base requirements for ITER TS diagnostics. The paper describes the diagnostic system that allows precise measurements of TS signals, as well as the results of the first measurements of electron temperature and density in both central region of the plasma column and scrape-off layer. The system provides measurements of electron temperature $T_e$ in the range of 5 eV to 5 keV and density $n_e$ in the range of $5 \cdot 10^{17} \div 3.25 \cdot 10^{20}$ $m^{-3}$. The use of two ITER-grade probing lasers of different wavelengths (Nd:YAG 1064.5 nm and Nd:YLF 1047.3 nm) allows reliable measurement of $T_e$ in multi-colour mode, i.e., assuming that spectral calibration is unknown.*

# Introduction

**T**homson **s**cattering (TS) is one of the most important diagnostic tools for plasma research in tokamaks and stellarators. In order to create an industrial thermonuclear power plant, a thorough research is required on such experimental facilities as ITER [1], JT-60SA [2], W7X [3], EAST [4], TRT [5] and other modern machines. As plasma parameters approach those of reactors, the requirements for diagnostic systems, in particular, for TS diagnostics, increase. Currently, the ITER project sets the highest standards, requiring measurements in stationary mode with a frequency of 100 Hz [6]. For the TS system developed for the ITER divertor, it is supposed to use a Nd:YAG laser (1064 nm) as the main probing one, providing scattering signals in near infrared region, where plasma radiation is significantly less than in the visible spectrum. The requirement for high measurement accuracy (the error of a single measurement should be <10% for electron temperature $T_e$ and <5% for density $n_e$) with intense background radiation [7] [8] [9] forces the use of short laser pulse ≤4 ns FWHM with a sharp leading edge. The best solution for recording signals, which is supposed to be implemented in all three ITER TS diagnostics, is high-speed (GHz) digitizers used to record the time trace of TS signals with a wide bandwidth. These digitizers provide an easy way to measure TS signals from multiple probing lasers of different wavelengths, synchronized on a scale of a few nanoseconds. Such a multi-colour approach determines $T_e$ even without spectral calibration [9] [10] [11]. The ability to operate without spectral calibration is crucial for reactor conditions, where a significant part of the optical elements will be unavailable for revision for a long time.

This paper is devoted to the recently developed TS diagnostic system for probing plasma of the Globus-M2 spherical tokamak in the equatorial plane. The Globus-M2 experimental program [12] is mainly focused on the study of plasma heating and non-inductive current-drive in a compact spherical tokamak with an outstanding toroidal magnetic field $B_T$ up to 1 T and a plasma current $I_p \leq 0.5$ MA. The measured values of $n_e$ are ranged from the extremely low values of $5·10^{17}$ m$^{-3}$ in the **S**crape **O**ff **L**ayer (SOL) to the reactor-like value ~$10^{20}$ m$^{-3}$ in the plasma core. The corresponding $T_e$ is expected to be from a few eV to a few keV. To fulfill the machine experimental program, both $T_e$ and $n_e$ profiles and their evolution during plasma discharges must be routinely measured. The measurements must be performed in the equatorial plane and should cover the space from SOL on the low-field side to at least the magnetic axis with a spatial resolution of 10-20 mm in projection onto the major radius $R$. The optical layout in the equatorial plane reduces the complexity of experimental data processing since an inward radial displacement of the plasma column does not interfere with temperature and density measurements on the magnetic axis. Such diagnostics layout, on the contrary, can be used to verify the position of the magnetic axis as well as the position of the last closed magnetic surface. Additional measurements on the high-field side are desirable since they allow evaluation of the toroidal rotation profile based on the asymmetry of the $n_e$ profile [13].

The developed TS diagnostics fulfills all these demands and satisfies the basic requirements for TS diagnostics in thermonuclear reactors [14]. These include high reliability in every single measurement, steady-state operation and compatibility with real-time data processing required for feedback control. The system prototypes the ITER divertor TS diagnostics utilizing a number of ITER-grade components and testing the multi-colour self-calibration technique. This spectral transmission calibration, which is necessary to eliminate errors due to spectral transmission distortion, is used by all three TS systems in ITER [6] [9] [11]. However, the multi-colour probing approach has never been tested on tokamaks with synchronized laser pulses at different wavelengths spaced within a time interval small enough to guarantee permanence of the local plasma parameters [15, 16].

The paper is organized as follows: 1 – description of the TS diagnostics layout at the tokamak, 2 – outline of the data acquisition system, 3 – analysis of the first results of $T_e$ and $n_e$ measurements, 4 – discussion of the experiments on multi-colour probing as a self-calibration technique.

# 1. Laser injection system and optical layout

The developed TS system consists of three main parts (see Figure 1): the beam injection system, the collection optics and the data acquisition system.

The beam injection system is based on two probing lasers. The main one is the Nd:YAG (1064.5 nm, 1-330 Hz, 0.1-3 J, 10 ns FWHM) pulsed laser (see Figure 2) manufactured by «Lasers & Optical Systems» Co. Ltd [17]. It is the LD-pumped Q-switched laser based on **M**aster **O**scillator + **P**ower **A**mplifier (MOPA) configuration. The beam divergence of 0.1 mrad just 1.3 times exceeds the diffraction limit for the exit aperture of 15 mm. The pulse energy is monitored by Ophir PE50-C pyroelectric energy meter in parallel with pulse shape recording. The laser can operate continuously for several hours, providing the diagnostics steady-state operation. Description of Nd:YLF laser used as an auxiliary one is presented in section 4 devoted to multi-colour measurements. The probing beams are directed from the separated laser room into the plasma via 6 mirrors over ~11 m (see Figure 1). Fraction of the probing beam, which was not reflected by the first mirror, is transmitted via optical fiber to the data acquisition system. This signal is used for synchronization, supplementary energy monitoring and pulse shape control for each laser pulse.

The single-pass probing is performed in the tokamak equatorial midplane with the beam dump installed outside the vacuum vessel. The layout of the collection optics is presented in Figure 3 demonstrating the equatorial cross-section of the Globus-M2 tokamak. Laser radiation is focused in the vacuum vessel of the tokamak using a single anti-reflection coated lens with focus length of 2.5 m and aperture diameter of 65 mm. The in-vessel baffles installed in the input beamline reduce intensity of the stray light. The observed probing chord corresponds with the tokamak major radius $R$ range from 23 to 56 cm ($r/a$ from -0.5 to 0.8).

The five-lens collection objective focuses scattered light onto fiber bundles. The lens magnification is in range from 4.1 on LFS to 2.8 on HFS, see Table 1, Figure 5. The optical scheme of the system is close to telecentric meaning the rays in the image-space are almost parallel to the optical axis. This greatly facilitates alignment of the fiber bundles. The system has 18 sockets for optical fiber bundles used to transmit the collected light to spectrometers. At the current stage of work, the diagnostics was equipped with 8 regular fiber bundles and 1 special fiber bundle with enhanced spatial resolution to study plasma in the vicinity of the separatrix (see Figures 4, 5). Each fiber bundle (CeramOptec GmbH [18]) consists of 188 fused silica 196/220 μm optical fibers with numerical aperture $NA$ of 0.22.

The fiber bundles on the tokamak-side are packed into 5.40×1.55 mm rectangles shown in Figure 4. The image size of the fiber bundle face on the probing beam line is defined by the lens magnification and determines spatial resolution in plasma. Figure 5 shows the projection of the scattering volumes on the tokamak poloidal plane and the plasma magnetic configuration typical for Globus-M2 with a single-null lower-divertor configuration. The low field side mid-plane separatrix position can be varied within the range $R_{sep}$ = 51-60 cm. In order to study steep gradients of $T_e$ and $n_e$, the spatial resolution at the plasma edge was increased using special fiber configuration (see Figure 4b): one fiber bundle is split into two halves, forming two spatial points for two spectrometers.

At this stage, TS measurements were performed on the low-field side at 10 spatial points located in the area from the magnetic axis to the separatrix. The remaining lens sockets are expected to be used for system expansion in future upgrades. The main parameters of the collection optics are listed in Table 1 for several spatial points. As it can be seen from Table 1 and Figure 3, the scattering angle increases for the peripheral points, which is beneficial for low $T_e$ measurements due to broadening of the backscattered TS spectrum.

The overall collection system transmittance can be estimated as 0.3, taking into account Fresnel reflections, the optical fiber packing as well as internal losses in fibers and lenses.

## 2. Spectroscopic and data acquisition systems

At the first stage the registration system consists of 10 polychromators. Each of them has 6 spectral channels based on band-pass filters. The polychromators have the same optical design previously used for TS diagnostics of the tokamak T15MD [19]. The set of filters, which is optimized for measuring $T_e$ in the range of 5-5000 eV, (see Table 2) is the same for all 10 polychromators. The TS spectra for different $T_e$ are shown in Figure 6 and compared with the transmission curves of the spectral channels.

Each spectral channel is equipped with Hamamatsu S11519-15 [20] **A**valanche **P**hoto**d**iode (APD) used with the internal gain $M = 100$. The APD gain thermal drift is compensated by a bias voltage regulation individually for each detector. The compensation system provides gain stabilization within ±0.5% in a wide temperature range from 17.5 to 35 °C [21]. The APD diameter of 1.5 mm was chosen to maximize the detector surface area, without limiting the detector bandwidth. The custom APD preamplifier circuit board [21], designed for low-noise operation, provides a wide bandwidth >250 MHz, reaching the bandwidth of the APD itself. The 2 MHz low-frequency cut-off of the amplifier bandwidth is used to suppress strong background light. The requirement to measure $n_e$ in the range from $5 \cdot 10^{17}$ to $3 \cdot 10^{20}$ m$^{-3}$ leads to a 600-fold difference in the expected TS signal amplitudes. To record signals in such a wide dynamic range, each signal channel is equipped with two parallel high-frequency outputs with gains of $G = 2$ and $G = 10$. Simultaneous parallel digitizing of these outputs provides accurate measurements of both high and low $n_e$, which is especially important in the regions with steep $n_e$ gradients.

All TS signals are recorded by CAEN V1743 digitizers, based on the SAMLONG chip with the 3.2 GS/sec sampling frequency at the 12-bit resolution. This is a second generation of the commercially available **S**witched-**C**apacitor **A**rray (SCA) digitizers [22], providing windowed multi-GHz sample rates at a relatively low cost. The digitized TS data is stored in central computer in real-time with a continuous event rate up to 1.4 kHz in full-window mode. This ensures reliable signal registration at the laser repetition rate of 330 Hz and allows an increase in the probing frequency. The internal memory of the digitizers allows measurements in burst-mode up to 8 kHz to study fast plasma processes. The data acquisition system is compatible with real-time data processing [23], which is in great demand for tokamaks with advanced plasma control.

The high sampling frequency improves SNR [19] and data reliability [24], compared to charge-sensitive integrators [25] used for TS applications in the past. The wide bandwidth of the detector and preamplifier, together with a short laser pulse, provide the pulse shape information for advanced signal processing [26]. In the case of a short probing laser pulse, the pulse shape analysis is useful for detection and rejection of non-TS signals such as stray light, dust scattering, high-energy quantum capture or over-scale events that can be mistaken for TS signals. The chosen digitizers provide a time step of 0.3125 ns, which is sufficient to record the auxiliary laser pulse shape at FWHM of 3 ns. The 320 ns window width is suitable for recording both the main and auxiliary laser pulses in the frame of one data page.

Additionally, each spectral channel is equipped with a low-pass output required for background signal registration for estimation of the measurement errors [21] as well as for spectral calibration. Signals of the low-pass outputs captured with 500 kHz 16-bit ADC are similar to passive broadband spectroscopy data and can also be used to measure the effective ion charge distribution [27].

The data acquisition system is placed in a separate electromagnetically shielded room. All 10 polychromators are housed in a single 19″ rack (Figure 7), including a VME crate for 8 CAEN V1743 boards providing 128 high-pass channels and 4 ADC boards for 64 low-pass channels. The remaining part of the crate is used by the accessory boards used to distribute the trigger pulse and to record the shape of the laser pulse. The central computer is fitted in the same rack. The modular approach ensured the compactness of the developed data acquisition system, ease of use and low sensitivity to electromagnetic interference.

## 3. First experimental results

The presented TS diagnostics is routinely used in plasma experiments since November 2020 [28]. Typical signal waveforms are shown in Figure 8. Each oscillogram corresponds to one spectral channel of polychromator for a single laser pulse. The TS signal is clearly observed at ~230 ns of the digitized window. The rest of the window is used to estimate the background noise level [21]. The presented data shows a remarkable signal-to-noise ratio even with the laser energy intentionally reduced by a factor of 4 and $n_e = 2.8 \cdot 10^{19}$ m$^{-3}$. The data processing algorithm is a classical one based on the weighted least squares fit [29]. An example of such a fit for the signals taken from Figure 8 is shown in Figure 9. The experimental data (bars) agree very well with the model spectrum (solid line) for the calculated $T_e$ and $n_e$ values.

An example evolution of local $T_e$ and $n_e$ values is shown in Figure 10. These measurements were taken during the Globus-M2 shot #39627 with neutral beam injection. The soft X-ray signal (see Figure 10.2) shows sawtooth oscillations during the flattop phase of the plasma current $I_p$, starting from 173 ms. Figure 11 shows a detailed scan of the profile evolution during an average sawtooth oscillation. These graphs show transition from the state of the mixed plasma core immediately after the sawtooth relaxation to the peaked profile before the next sawtooth relaxation. The new TS diagnostics makes it possible to localize the mixing radius $R_{mix}$ =49 cm equal to $r/a$ = 0.44. The equilibrium was studied using the ASTRA code. The code predicted safety factor $q = 1$ at $r/a = 0.4$ which is in a good agreement with the measured one.

Information about the outermost closed magnetic surface can be obtained from current filament reconstructions [30]. The ASTRA approximation [31] eq.88 is used to determine the internal magnetic poloidal flux structure. Assuming $T_e$ and $n_e$ being constant along a magnetic flux surface, one can estimate the stored electron energy $W_e$. The obtained $W_e$ showed good agreement with both ASTRA and PET [32] modeling. The described algorithm does not require a resource-demanding plasma equilibrium computation, making it possible to carry out express analysis of experimental results. Figure 10 shows evolution of the stored electron energy $W_e$. The obtained 2-dimentional poloidal distribution of $n_e$ can also be used to estimate the linear density $nl$. Comparison of the $nl$ estimated from the TS data with the data of microwave interferometer shows good agreement between these two diagnostics.

## 3.1 SOL measurements

High sensitivity of the new TS diagnostics allows measurement of low $T_e$ and $n_e$ inherent in plasmas of **S**crape-**O**ff **L**ayer (SOL). In Figure 12, the values of $T_e$ are plotted against distance to separatrix. The separatrix position was determined by the current filament method from magnetic diagnostics data. In the experiment, the plasma separatrix was moved away from the outer wall during stationary phase of the tokamak discharges №39586, №39589 and №39590. This allowed scanning the peripheral region of plasma with the fixed TS lines of sight. The experimental data were obtained with intentional decrease in the energy of the probing laser by a factor of 4 and halved optical fiber bundles, resulting in overall decrease in TS signal by a factor of 8. Even with the significantly reduced performance, the new TS system provided reliable measurements far beyond the separatrix, with the value of $n_e$ at the separatrix of $1 \cdot 10^{19}$ m$^{-3}$. Registration of the $T_e < 10$ (eV) makes it possible to estimate the exponential decay length of $T_e$ behind the separatrix. A separate article will be devoted to a detailed comparison of the $T_e$ and $n_e$ in SOL measured by TS with these measured by Langmuir probes.

TS measurements in SOL demonstrated the lower limit of the diagnostics dynamic range. Figure 13 shows an estimate of $T_e$ error in the experiments with a significant decrease in the TS signal: the probing laser energy was 0.25 of the maximum pulse energy for this laser, and the scattering length was halved. The upper limit of measurable $n_e$ is determined by overloading of the amplifier and digitizer (see Figure 13). The resulting measurement range of $n_e$ is from $5 \cdot 10^{17}$ to $3.25 \cdot 10^{20}$ m$^{-3}$ for the system at full performance. The wide dynamic range of the developed polychromators enables the use of the same instrument for measurements in all areas of the Globus M2 plasmas from the core to SOL. The high sensitivity of the system made it possible to reduce the amount of collected light and thereby improve the spatial resolution to 2.5 mm near separatrix.

## 4. Multi-colour TS measurements, experimental conditions and preliminary analysis

Spectral characteristics of a collection system in reactor conditions changes over time and is generally unknown [6]. TS system with multiple lasers operating at different wavelengths, referred to as multi-colour TS, can provide in-situ calibration of relative spectral sensitivity [10] [11]. Multi-colour TS approach requires that the local plasma parameters remain unchanged in the time interval between probing pulses with different wavelengths. To ensure this condition for a turbulent edge plasma, probing lasers must be synchronized on the scale of tens of nanoseconds. With this requirement met, the $T_e$ can be determined by minimizing the sum of squared deviations [10]:

$$\sum_{i=1}^{N} \frac{(S_i - \gamma \cdot F_i(T_e))^2}{\sigma_{S_i}^2} \to min, \qquad (1)$$

where $i$ marks spectral channel; $S_i$ is the ratio of TS signals from lasers with wavelengths $\lambda_{01}$ and $\lambda_{02}$ recorded in the same spectral channel: $S_i = U_{\lambda_{01 i}}^{TS}/U_{\lambda_{02 i}}^{TS}$; $\gamma$ is the ratio of probing pulse energies at $\lambda_{01}$ and $\lambda_{02}$; $\sigma_{S_i}^2$ is an estimate of the $S_i$ variance; and $F_i(T_e)$ is the expected ratio of TS signals depending on $T_e$. The spectral channel sensitivity is cancelled as it occurs in both numerator and denominator of $F_i$:

$$F_i(T_e) \approx \frac{\int_{\lambda_{min_i}}^{\lambda_{max_i}} \sigma_{TS}(T_e, \lambda, \lambda_{01}) \cdot d\lambda}{\int_{\lambda_{min_i}}^{\lambda_{max_i}} \sigma_{TS}(T_e, \lambda, \lambda_{02}) \cdot d\lambda} \qquad (2)$$

Here $\lambda_{min_i}$ and $\lambda_{max_i}$ are the boundary wavelengths of the corresponding spectral channel, $\sigma_{TS}(T_e, \lambda, \lambda_0)$ is TS scattering power density.

The multi-colour TS approach was tested in plasma experiment on the upgraded TS diagnostics of the Globus-M2 tokamak [33]. The main Nd:YAG 1064.5 nm laser was complimented with an auxiliary Nd:YLF laser operating at 1047.3 nm. This laser was developed as a calibration laser for the ITER divertor TS diagnostics [34] and is supposed to be used for spectral calibration in a relatively narrow wavelength range of 150 nm corresponding to $T_e$ < 1000 eV [9]. The auxiliary laser pulse duration was 3 ns FWHM, and the pulse repetition rate was 50 Hz. The Q-switches of the main and the auxiliary lasers were synchronized with a relative time delay of ~60 ns. The operating frequency of the main laser was reduced from 330 to 300 Hz, so that every sixth pulse of the main laser was accompanied by a pulse of the auxiliary laser. The use of wide-bandwidth detectors together with short time delay between pulses made it possible to record scattering signals from both lasers within one time window of the digitizer.

Spectral sensitivities of the polychromator channels are shown in Figure 14 together with the TS spectra for both probing wavelengths. Since TS polychromators of the Globus-M2 tokamak were not designed for operation with 1047.3 nm laser, filter of the second spectral channel (see Table 2) has not enough contrast against this wavelength. Thus, the second spectral channel in these experiments was turned off in all polychromators.

Figure 15 shows the expected ratio of TS signals in channels of the spectrometer $F_i(T_e)$ see (2), for probing wavelengths $\lambda_{01}$ = 1064.5 nm, $\lambda_{02}$ = 1047.3 nm. The value of $F_1(T_e)$ is practically independent of $T_e$, which is typical for the vicinity of the critical wavelength calculated as $\lambda_{crit} = \sqrt{\lambda_{01} \cdot \lambda_{02}}$ = 1056 nm (see eq.8 [10]), [35] – the wavelength, where the ratio of multicolour TS signals depend only on probing energy relation. The measurement of low $T_e$ <50 eV was significantly complicated by the turned off second spectral channel, described above. The measurement of $T_e$ > 400–500 eV from the ratio of TS signals at the selected two probing wavelengths is also limited, since the functions $F_3(T_e)$ and $F_4(T_e)$ weakly depend on $T_e$ in this range (see Figure 15). Despite the fact that $F_5(T_e)$ strongly depend on $T_e$ for $T_e$ < 1000 eV, the measurement accuracy of the TS signal ratio $S_5$ (see Figure 14) is low for $T_e$ in this region. The resulting estimations of the expected errors in measuring both $T_e$ and $\gamma$ by the multi-colour technique are shown in Figure 16. Taking into account all the described limitations, the expected error in measuring $T_e$ by the multi-colour method is of 15-20% for $T_e$ in the range of 50-250 eV. The expected measurement error of multi-colour TS with unknown spectral calibration is 3-4 times higher than of the traditional TS approach with 1064 nm probing and known spectral calibration. However, the accurate measurement of $\gamma$ is possible in a wide $T_e$ range (see Figure 16), as $S_1(T_e)$ nearly linear depend on $\gamma$.

*4.1 Multi-colour TS measurements, experimental results*

Figure 17 shows the main plasma parameters of discharge #40204 with multi-colour TS experiment and the example oscillograms of synchronized multicolour TS signals. The first pulse corresponds to the main 1064.5 nm laser, the second pulse to the auxiliary 1047.3 nm laser with ~60 ns delay between pulses. The $T_e$ profiles measured by the classical TS method with each laser are shown in Figure 18 for 3 different phases of the tokamak discharge. Changes in the local $T_e$ value that occur between the main and auxiliary laser pulses do not exceed the measurement error. In this experiment the auxiliary laser energy was 2-3 times lower, than the main one. This results in lower measurement accuracy with the auxiliary laser, than with the main one.

The sample of $T_e$ values at all spatial points was formed from several tokamak discharges. All measurements with $T_e^{1064}$ > 1 keV were removed from the sample, where $T_e^{1064}$ is the electron temperature determined by the classical TS method with the main 1064.5 nm laser. This filtering was performed due to the limitations, described in the beginning of the section 4. The obtained sample is shown in Figure 19 a, which compares the $T_e$ calculated by the classical approach for each laser: $T_e^{1064}$ and $T_e^{1047}$. Figure 19 b shows the distribution of the deviation between $T_e^{1064}$ and $T_e^{1047}$ (the function ($T_e^{1047}/T_e^{1064}-1$)), which is close to the normal one, has no systematic shift and the average deviation is at the level of 10%

The multi-colour TS approach without using the spectral calibration data provides $T_e^{multicolour}$ value by minimizing the expression (1). Figure 20 shows the resulting distribution ($T_e^{mutilcolour}/T_e^{1064}-1$), which slightly differs from the normal one, having a small "tail" in the high $T_e$ region. The standard deviation of the $T_e^{mutilcolour}$ from $T_e^{1064}$ is of 22%, with a mean value shift of ~8%. The deviation of the obtained distribution between $T_e^{multicolour}$ and $T_e^{1064}$ from the normal one is caused mainly by the points corresponding to high $T_e$.

To prove this assumption, the following numerical experiment was carried out. At the first step, the "true" values $T_e^{true}$ and $n_e^{true}$ were set. Then, the expected average TS signals were calculated for the polychromator spectral channels $i$ = 1, 3, 4, 5 for the probing wavelengths $\lambda_{01}$ = 1064.5 nm and $\lambda_{01}$ = 1047.3 nm ($U_{\lambda_{01 i}}^{TS}$, $U_{\lambda_{02 i}}^{TS}$), as well as signal variances $\sigma_{\lambda_{01 i}}^{TS}$ and $\sigma_{\lambda_{02 i}}^{TS}$ were estimated. Based on the obtained values, a synthetic sample of the "measured" scattering signals was generated, assuming their normal distribution. Finally, the $T_e^{synth}$ parameter was determined by minimizing expression (1) using the least squares fit (see Figure 21) for each point of the obtained data set. It can be seen, that for moderate $T_e^{true}$= 165 eV, the distribution of $T_e^{synth}$ is normal without significant shift in the mean value. For $T_e^{true} = 500\ eV$, there is a deviation from the normal distribution of $T_e^{synth}$, as well as a difference between the mean value $T_e^{synth}$ and the value $T_e^{true}$, which was observed in the plasma experiment analysis above.

# 5. Conclusions

The recently developed Thomson scattering diagnostics was put into operation at the Globus-M2 tokamak. The developed system has shown high performance in plasma experiments and now is used as a monitoring tool. The pulsed steady-state LD-pumped probing laser with energy tunable in the range of 0.1-3 J provides measurements with repetition rate up to 330 Hz. The wide-bandwidth detectors (0.2 GHz) and ADC sampling rate as high as 3.2 GS/s allow measuring time traces of TS signal when probing with 3 ns pulses. The system allows measurement of $n_e$ in the range of $5 \cdot 10^{17}$ to $3.25 \cdot 10^{20}$ m$^{-3}$ and $T_e$ in the range of 5 to 5000 eV, providing reliable measurements both in plasma core and in the scrape-off layer. The developed acquisition system is fully compatible with multi-colour TS technique, that assumes plasma probing by laser pulses with a delay at a scale of a few ns. The first successful experiment on multi-colour TS with ns-scale delay between probing pulses was carried out using the new diagnostic system. The use of ITER-grade Nd:YAG 1064.5 nm and Nd:YLF 1047.3 nm probing lasers made it possible to perform $T_e$ measurements in the range of 30 to 1000 eV with an accuracy of 22% with unknown spectral calibration and despite the low energies of the probing pulses: 1 J and 0.5 J, respectively. The obtained accuracy of $T_e$ measurement by multi-colour method corresponds to the expected values. The developed system meets the basic requirements for a reactor and advanced plasma control applications.

Acknowledgements:

The experiments were carried out at the Unique Scientific Facility "Spherical Tokamak Globus-M", which is incorporated in the Federal Joint Research Center "Material science and characterization in advanced technology". The TS diagnostic system was deployed and put into operation (chapters 1 and 2) with financial support by RSF research project №17-72-20076. The plasma heating experiments, presented in chapter 3, were supported by Ioffe Institute (Russian Federation state funding assignments 0040-2019-0023). Preparations of multi-colour experiments, presented in chapter 4.1, were supported by Ioffe Institute (Russian Federation state funding assignments 0034-2019-0001). The development of the auxiliary Nd:YLF 1047 nm laser source presented in chapter 4 involved financial support by Ministry of Education and Science of the Russian Federation as part of grant № 17706413348210001850/29-21/01.

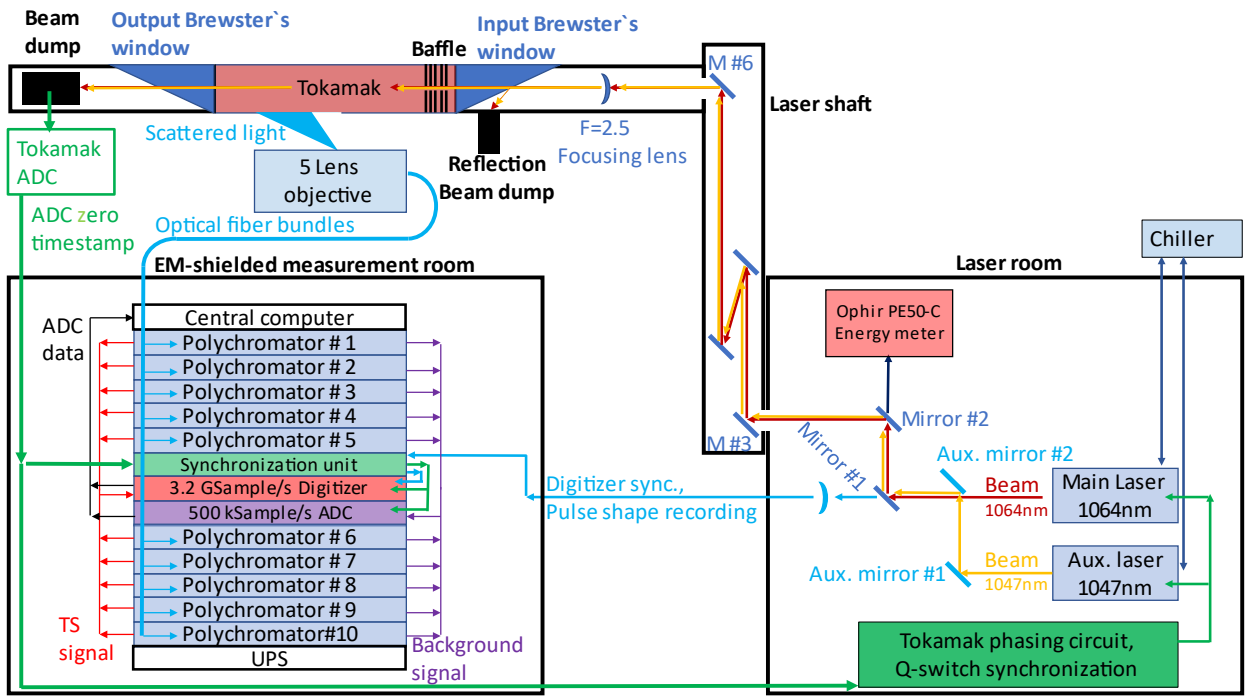

*Figure 1 TS diagnostics layout on the Globus-M2 tokamak. Lasers are placed in the room shown on the right. The paths of the laser beams are combined and launched via laser shaft into the tokamak, shown at the top. Scattered light is collected by the 5-lens objective and transmitted via fiber bundles to the measurement room, shown on the left. Each spatial point is processed by own polychromator with the use of external digitizers.*

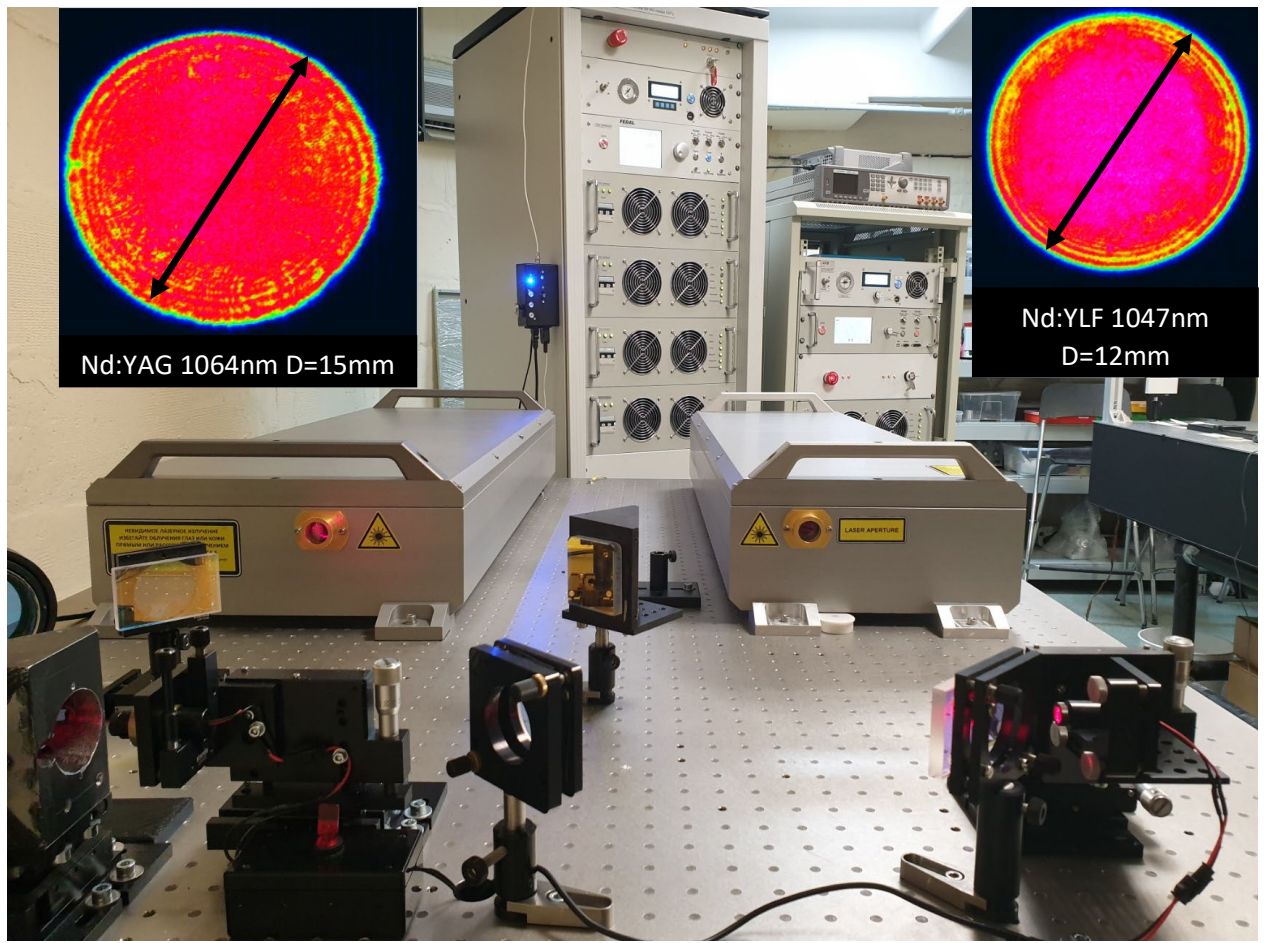

*Figure 2 The main probing laser – Nd:YAG 1064.5 nm 330 Hz 3 J 10 ns is on the left and the auxiliary laser used for multicolour calibration Nd:YLF 1047.3 nm 50 Hz 2 J 3 ns is on the right. Mirrors for beam paths combining are shown in the foreground. The controllers, power supply and cooling systems are housed into 19'' racks in the background. The near field energy distributions for both lasers are shown as insets.*

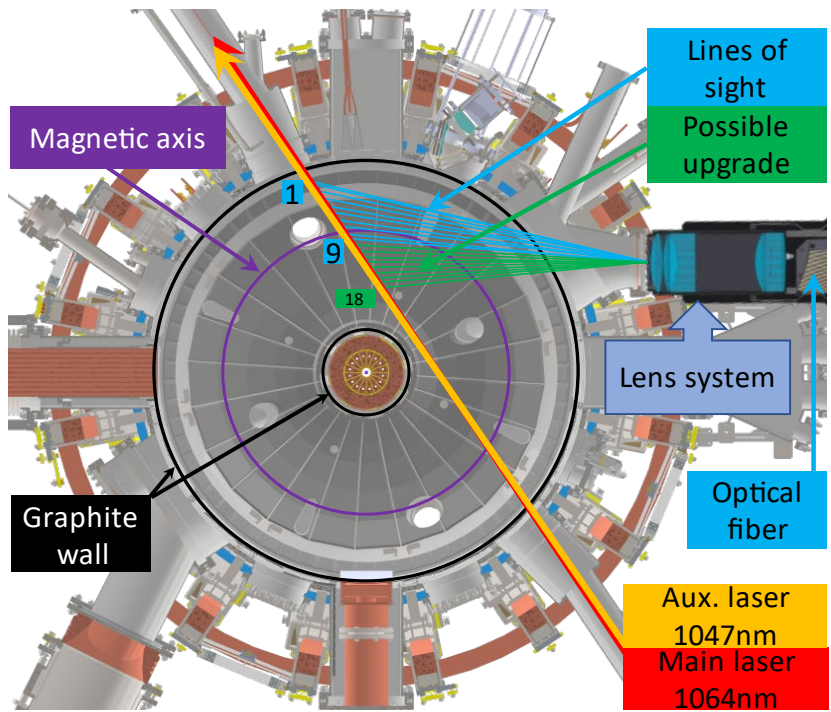

*Figure 3 The optical layout – an equatorial cross section view. The red and orange arrows mark the probing beams of the main and auxiliary lasers. The blue lines show lines of sight, used on this stage. The green lines show lines of sight that were not used now and are planned to be equipped with instruments with the diagnostics upgrade. The numbers correspond to the fiber bundle numbers (see Table 1). The purple circle indicates position of the magnetic axis, and the black circles indicate the first wall.*

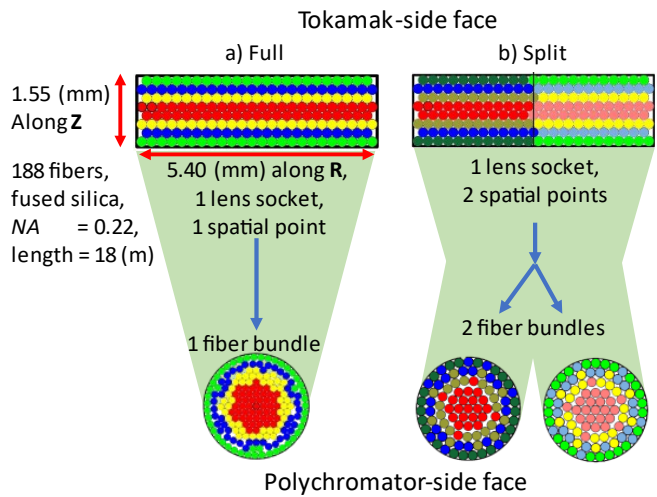

*Figure 4 The optical fiber package on the tokamak side (top) and at the polychromator side (bottom). The fibers (188 fibers in one bundle) are packed so that the image from the midplane is transferred to the center of the polychromator entrance. (a) common fiber bundle, (b) fiber bundle with splitting fibers into 2 groups providing enhanced spatial resolution.*

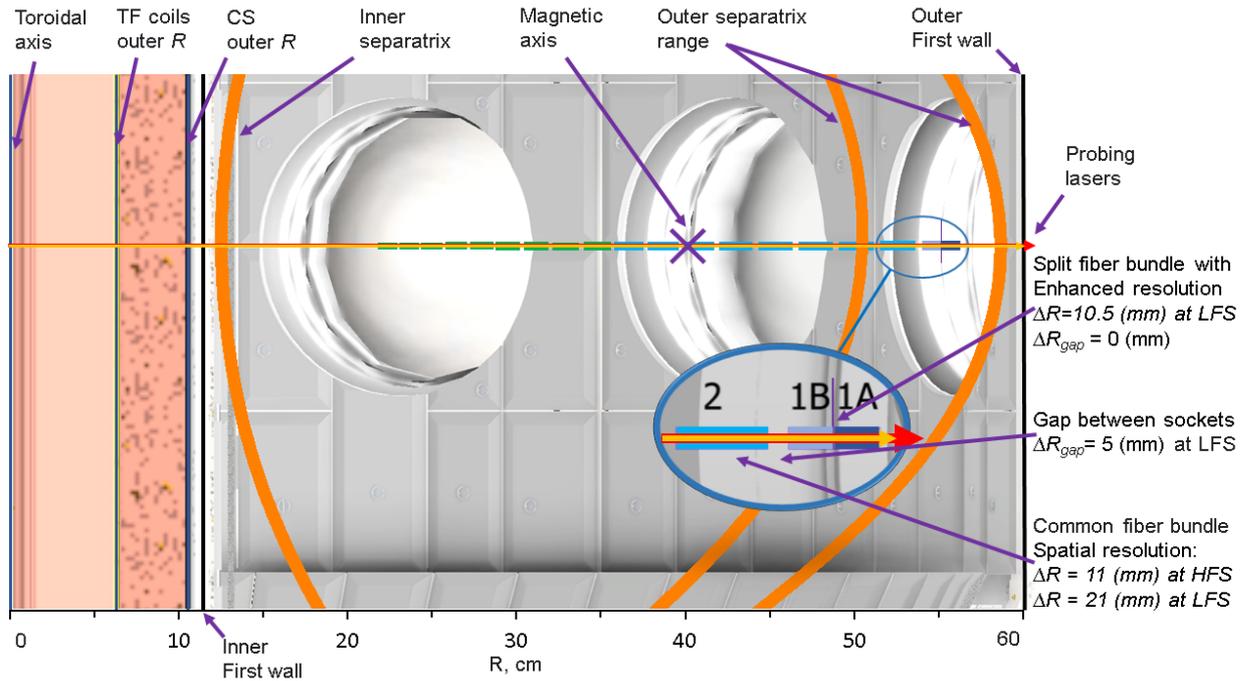

*Figure 5 Poloidal cross section of Globus-M2 with **R** scale shown at the bottom. Horizontal red and orange arrows mark probing chords for both lasers. The rectangles on the probing chord are poloidal projections of fiber bundles: blue – used in this work, green – expected to be used after update. The inset figure shows in detail the edge fiber bundles with enhanced spatial resolution. The split fiber bundle has no gap $\Delta R_{gap}$ between spatial points and provides twice improved spatial resolution $\Delta R$. The purple cross corresponds to approximate position of the magnetic axis. Orange curves show possible separatrix positions: the outer separatrix radius in equatorial plane can be varied in the range of 51-60 cm.*

| Fiber # | R, cm | $r/a_0$ | f-number | Scattering angle, ° | Lens magnification | Scattering length, mm | Spatial resolution, mm |
|---|---|---|---|---|---|---|---|
| 1 | 55.5 | 0.81 | f/7.4 | 137 | 4.1 | 22(11*) | 21(10*) |
| 5 | 45.8 | 0.41 | f/6.8 | 132 | 3.9 | 21 | 19.5 |
| 9 | 37.1 | 0.05 | f/6.4 | 127 | 3.4 | 18.5 | 16.5 |
| 14 | 28.5 | 0.31 | f/6.0 | 123 | 3.0 | 16.2 | 13.5 |
| 18 | 22.9 | 0.55 | f/5.7 | 118 | 2.8 | 15 | 11.0 |

*Table 1 Optical parameters for several spatial points are listed from LFS to HFS. Values marked with * are for the fiber bundle with enhanced spatial resolution.*

| Spectral channel | Channel center wavelength, nm | Channel width FWHM, nm |
|---|---|---|
| 1 | 1056.3 | 12.5 |
| 2 | 1039.3 | 21.5 |
| 3 | 1007.3 | 42.5 |
| 4 | 944.0 | 84.0 |
| 5 | 818.3 | 167.5 |

*Table 2 Polychromators spectral channels used for TS measurements.*

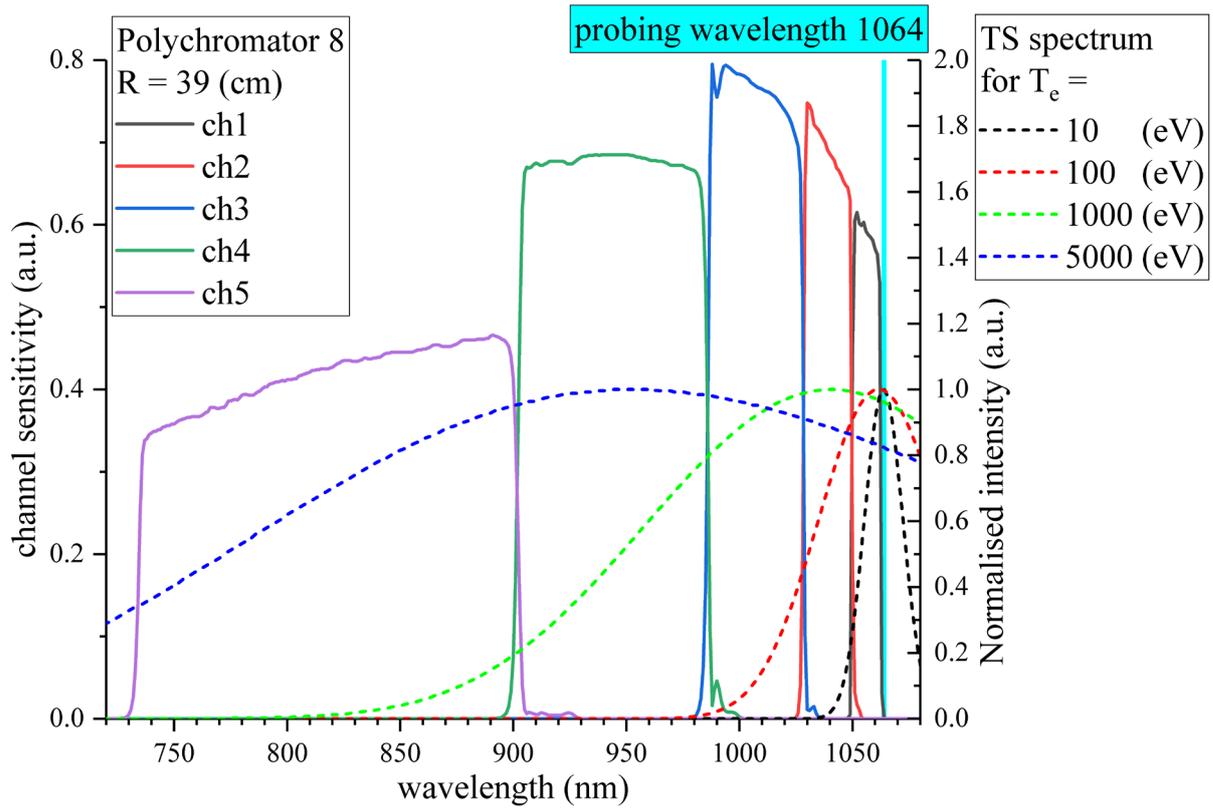

*Figure 6 Spectral sensitivity of the polychromator (solid lines scaled on the left side) compared to the normalized TS spectra at different $T_e$ (dashed lines scaled on the right side). The spectral sensitivity takes into account transmission of filters, sensitivity of detectors, and transmission from channel-to-channel inside the polychromator. The cyan vertical line marks 1064.5 nm – the main probing wavelength.*

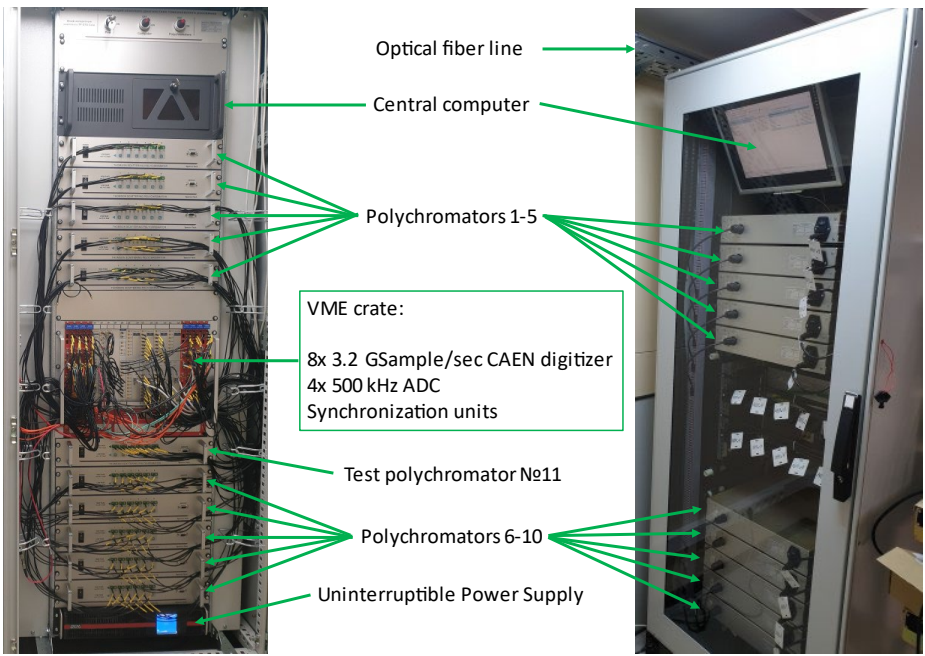

*Figure 7 The data acquisition setup. The front panel is shown on the left, the back one – on the right. The rack is organized as follows (from top to bottom): the optical fiber input, the central computer, 5 polychromators, VME crate, 5+1 polychromators, UPS. The VME crate is filled by (from left to right): 4 CAEN V1743 boards, 4 synchronization units, 4 500 kHz custom ADC boards, 4 CAEN V1743 boards. The 11$^{th}$ polychromator shown on the left is a test device added to the main system.*

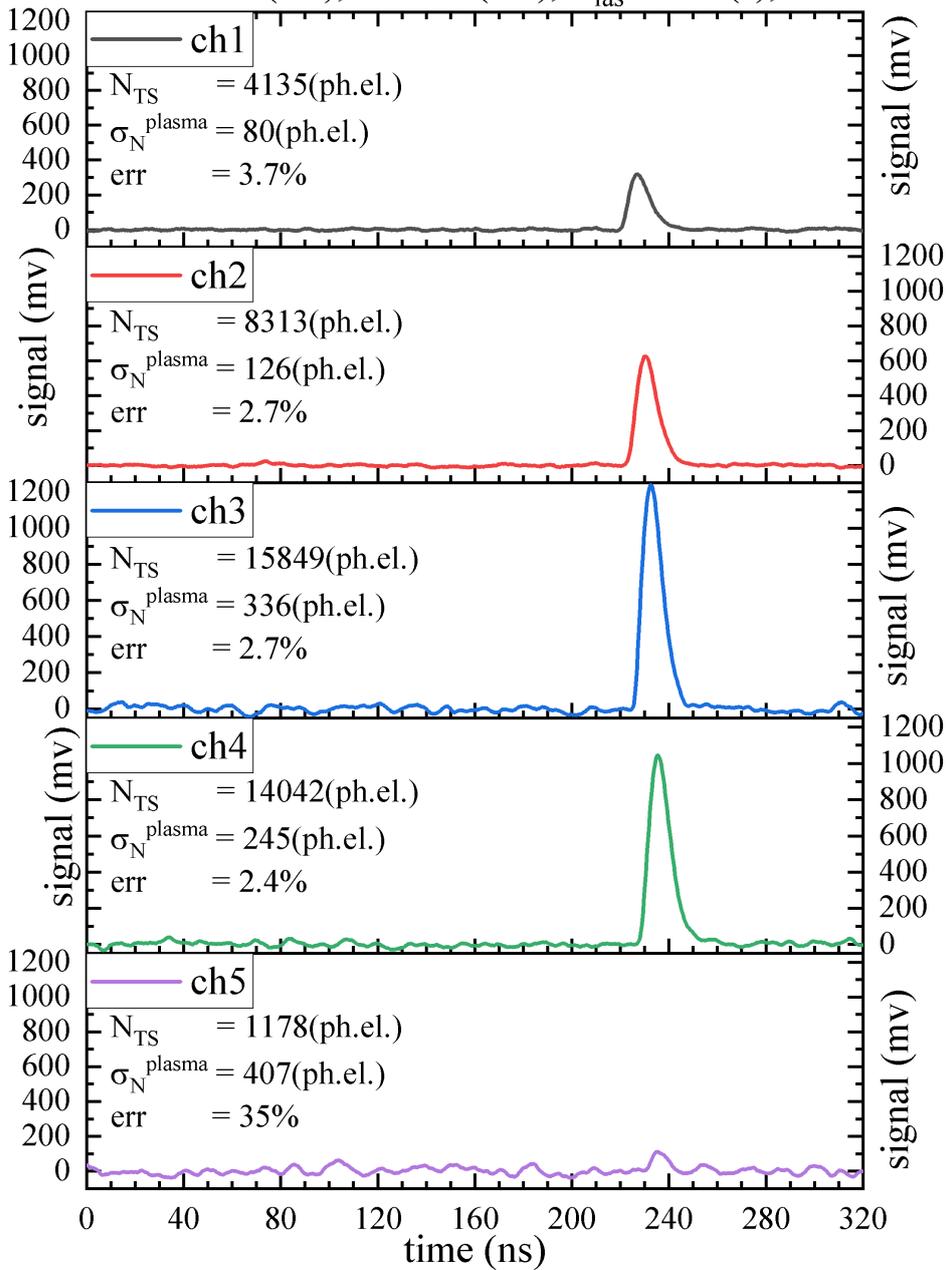

Figure 8 Time page of the fast digitizer for a single laser pulse registered at 203.6 ms of the Globus-M2 discharge #39627. Oscillograms of the TS signals were measured at the plasma core with **R** = 39 cm. Each spectral channel is plotted on its own diagram, including the calculated value of TS photoelectrons $N_{TS}$, the noise from the plasma background $\sigma_N^{plasma}$ and the estimated errors **err** of the TS signals. The probing laser energy was deliberately reduced to 0.7 J, $n_e = 3.6 \cdot 10^{19}$ $m^{-3}$, the signal was recorded from outputs with low-gain **G** = 2.

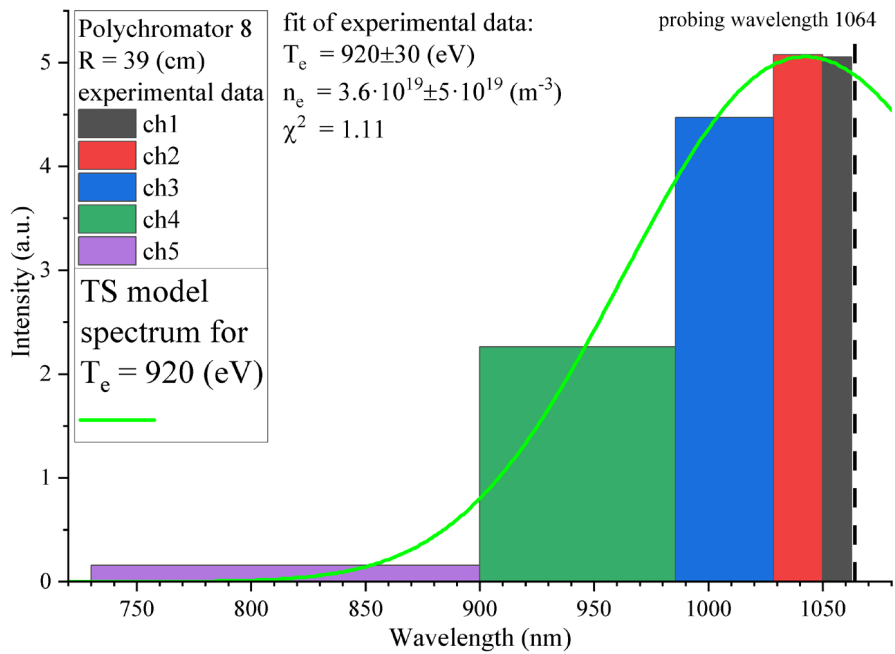

*Figure 9 Example of fitting signals (bars) from Figure 8 to the model TS spectrum (green line) for **$T_e$** = 920 eV. The fitting shows a good accordance of the model: the corresponding value of $\chi^2$ is of 1.11.*

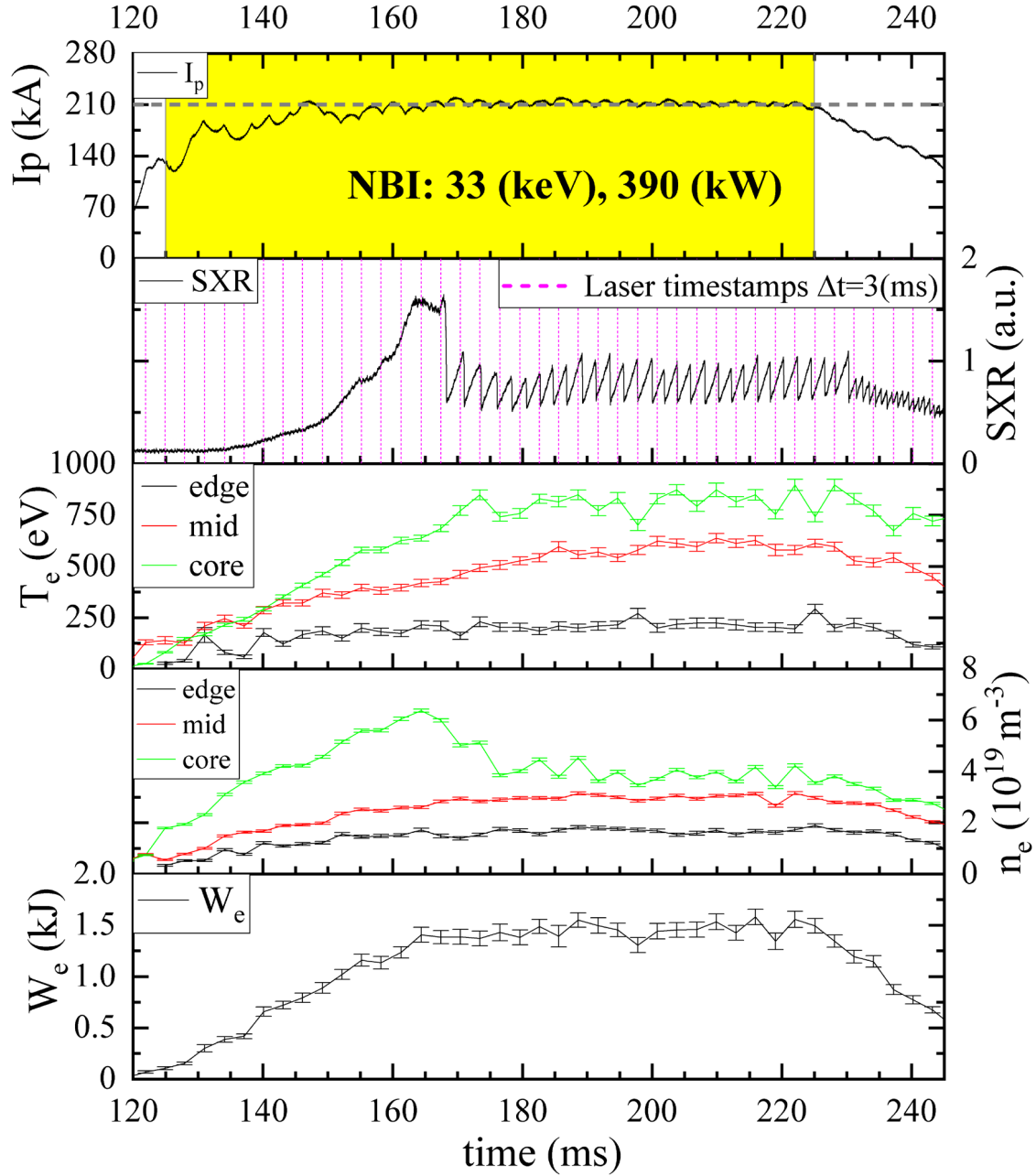

*Figure 10 Behavior of the main plasma parameters during discharge #39627 in the presence of sawtooth oscillations. From top to bottom: plasma current $I_p$ and neutral beam injection time, soft X-ray intensity and time marks corresponding to TS measurements, $T_e$ and $n_e$ in the center (core), at the edge (edge) and in the middle (mid) of the plasma column, $W_e$ – the stored energy in electrons calculated in the same spatial points.*

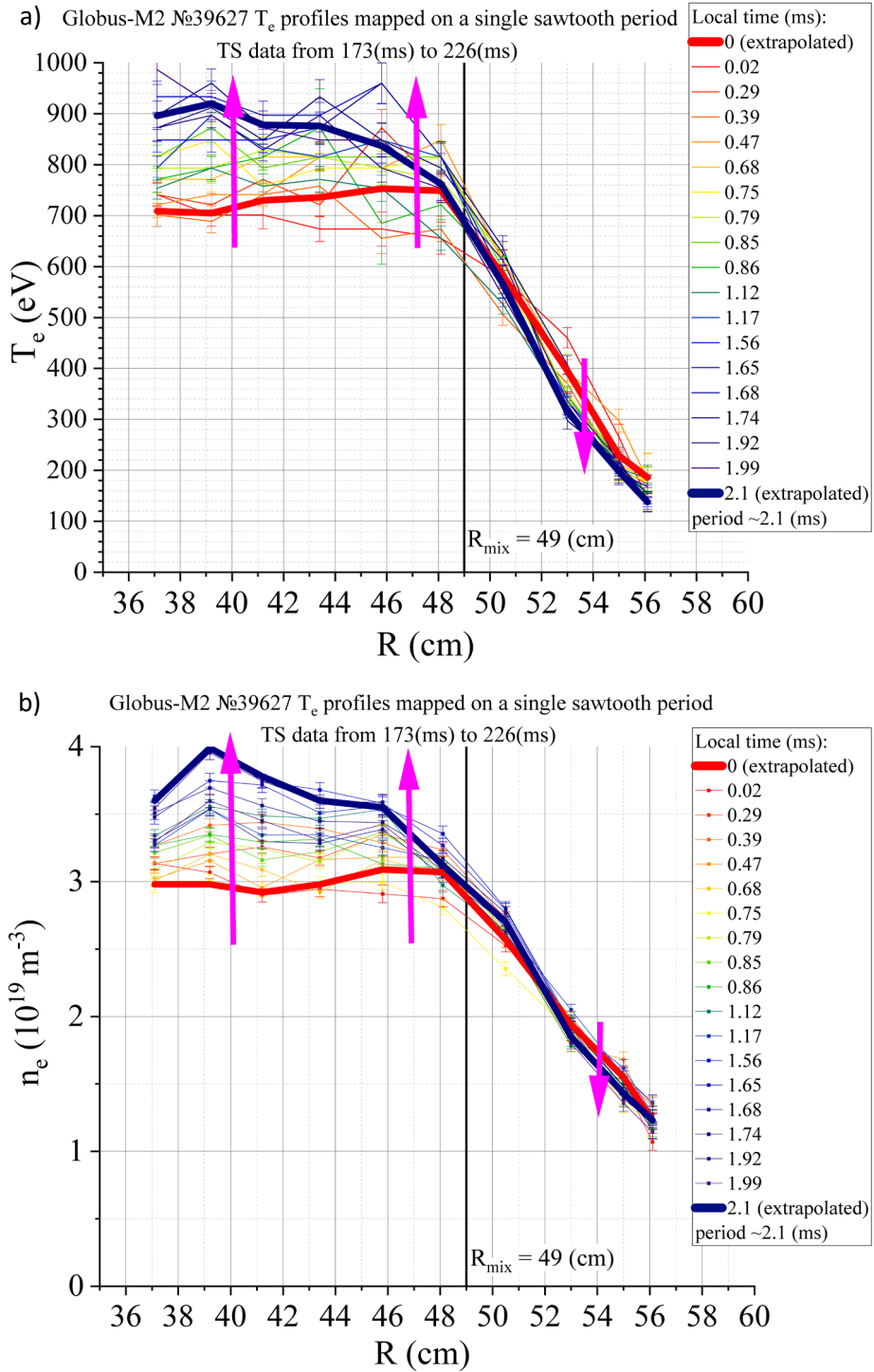

*Figure 11 Profile evolution during sawtooth oscillations in discharge #39627:(a) $T_e$, (b) $n_e$. Time is measured from the sawtooth relaxation, the profile colour changing from blue for the earliest measurements to red for the latest. The magenta arrows show the main trends in profile shape changes. The averaged mixing radius is $R_{mix}$ = 49 cm, equal to r/a = 0.44.*

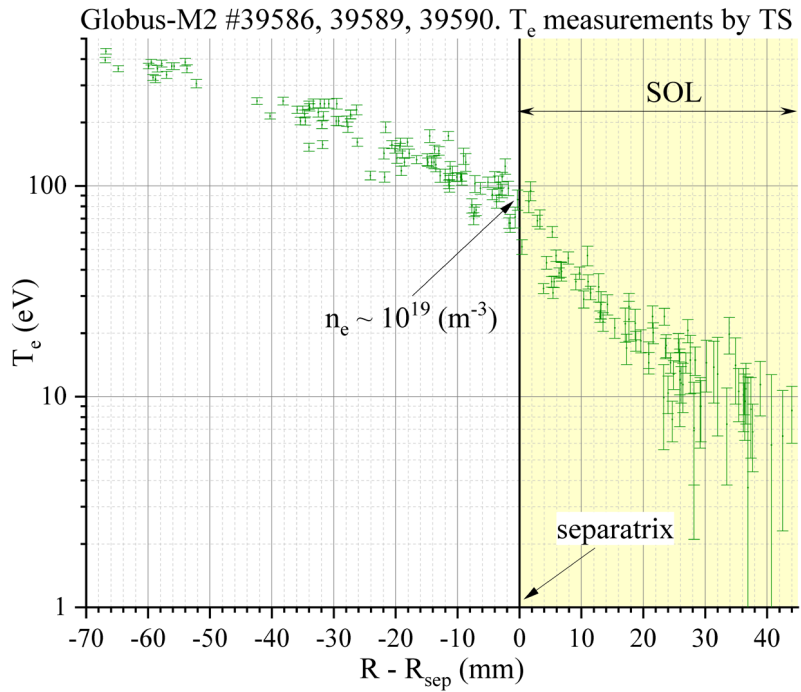

*Figure 12 Edge **T**$_e$ profile (log scale), measured by the equatorial TS system. The separatrix position moved during flattop of the discharges. The horizontal axis shows distance from the separatrix, calculated at the moment corresponding to the TS measurement. **n**$_e$ at the separatrix was of $10^{19}$ m$^{-3}$ and the reduced laser energy was of 0.7 J.*

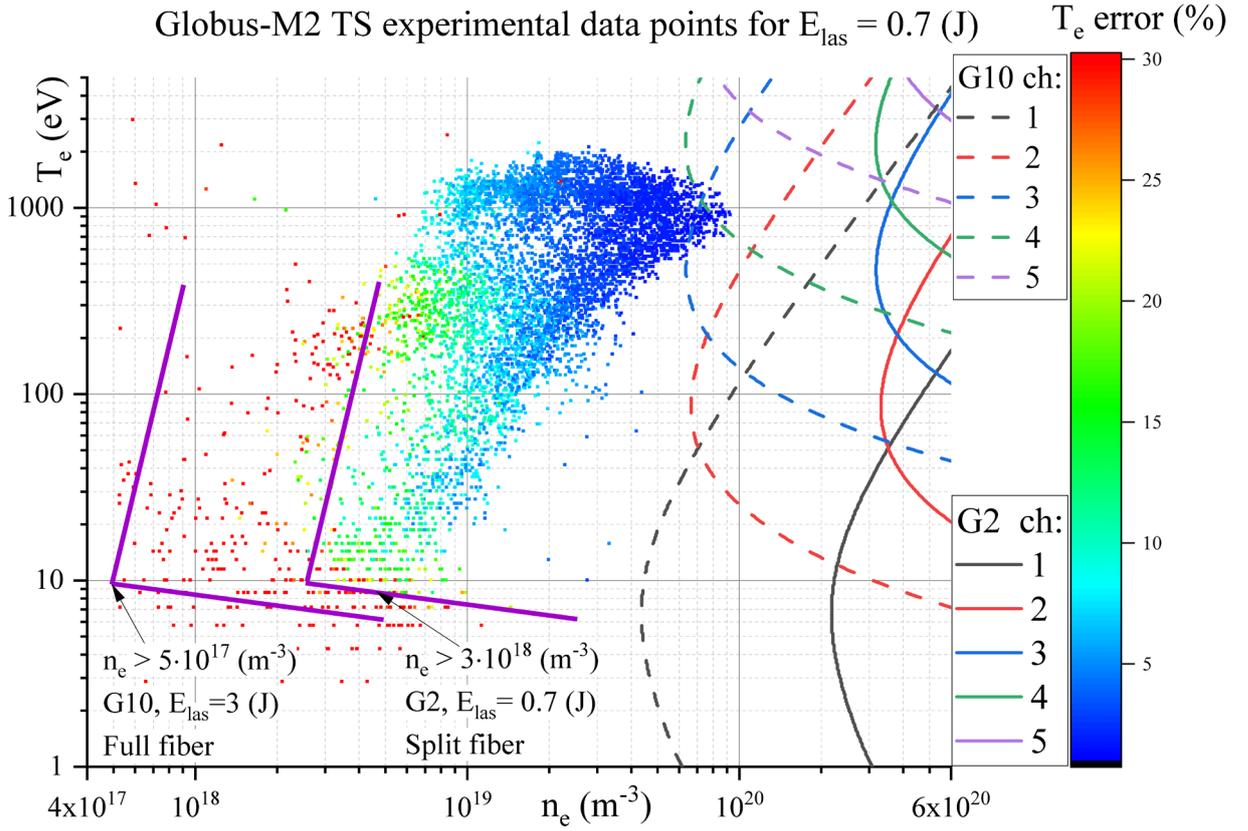

*Figure 13 Colourmap of the electron temperature $T_e$ measurement error $T_e$ error depending on $T_e$ and electron density $n_e$. Points represent experimental data, obtained with significantly reduced TS signals: laser energy $E_{las}$ = 0.7 J, channel amplification is $G$ = 2 and the fiber bundle size (scattering length) is halved. Under these conditions diagnostics allows reliable measurement for $n_e \geq 3 \cdot 10^{18}$ $m^{-3}$. In full performance the limit is $n_e \geq 5 \cdot 10^{17}$ $m^{-3}$. The high $n_e$ limit for $E_{las}$ = 0.7 J is determined by amplifier and digitizer overload, shown by dashed lines for each channel at $G$ = 10 and by solid lines for $G$ = 2.*

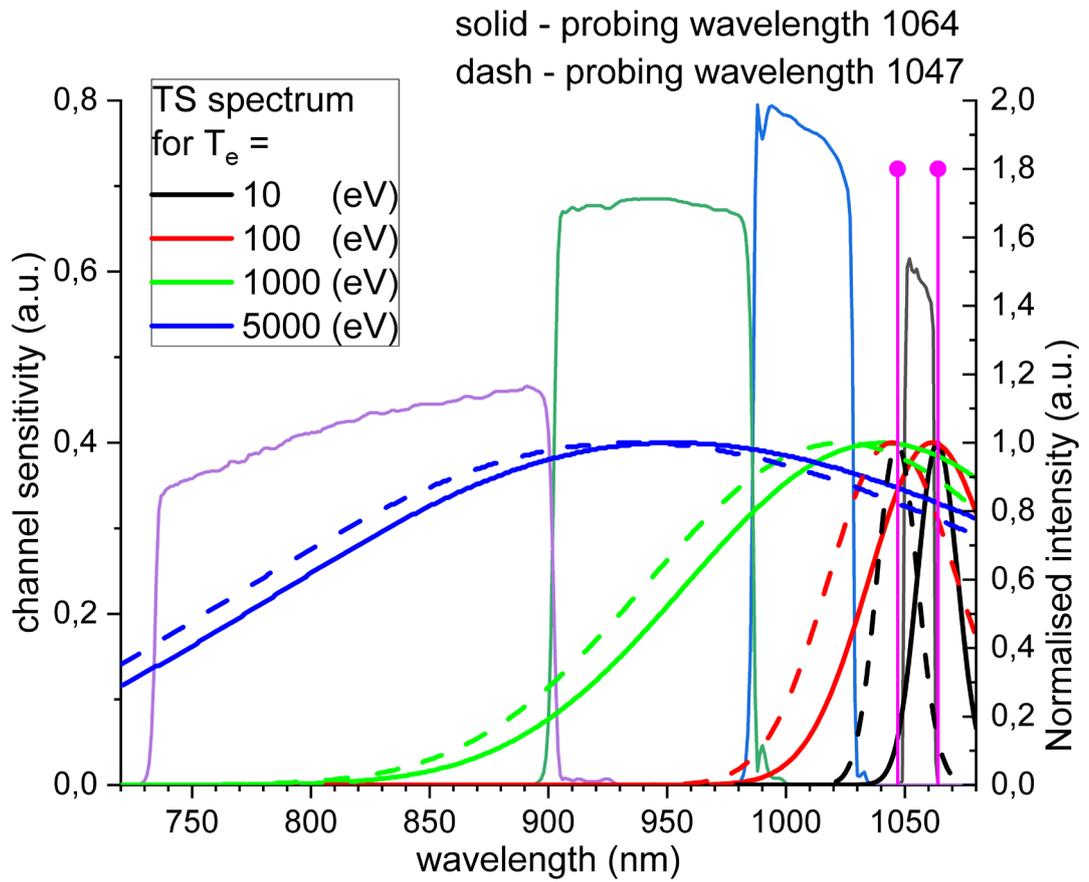

*Figure 14. Spectral sensitivity of the polychromator (see Figure 6) compared to the TS spectra for two probing wavelengths at different electron temperatures. The solid curve corresponds to the normalized scattering contours for $\lambda_{01}$=1064.5 nm, the dash curve for $\lambda_{02}$=1047.3 nm. The positions of the probing wavelengths are indicated by vertical magenta markers.*

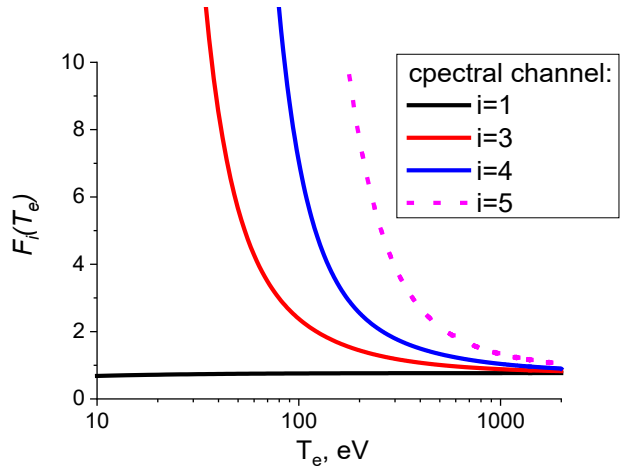

*Figure 15. The dependence of TS signals ratio in the selected polychromator channels, $F_i(T_e)$ see (2), for the probing wavelengths $\lambda_{01}=1064.5$ nm and $\lambda_{02}=1047.3$ nm. $F_1(T_e)$ is practically constant as the 1$^{st}$ spectral channel is thin and includes $\lambda_{crit}$. The $F_3(T_e)$ and $F_4(T_e)$ have strong dependency on $T_e$ in the range of 50-400 eV. $F_5(T_e)$ strongly depend on $T_e$ for $T_e < 1000$ eV, but the measurement accuracy in this region is too low. The proximity of the probing wavelengths results in weak $F_i$ dependency on $T_e > 1000$ eV.*

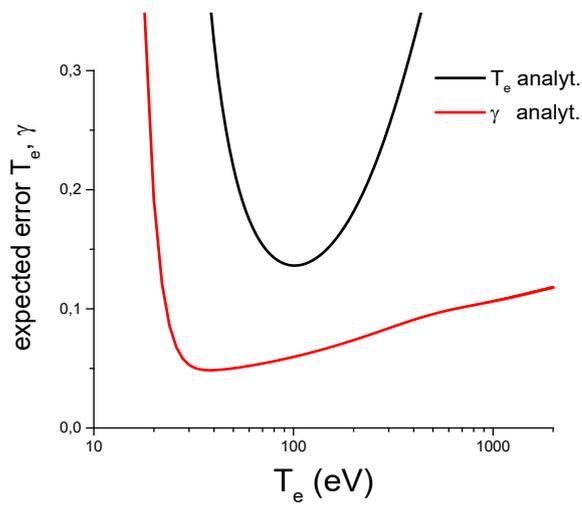

*Figure 16. Expected measurement error of electron temperature $T_e$ and $\gamma$ (see eq.1) by the multi-colour technique for the Globus-M2 TS collection system using probing wavelengths $\lambda_{01}=1064.5$ and $\lambda_{02}=1047.3$ nm with turned-off 2$^{nd}$ spectral channel of the polychromator.*

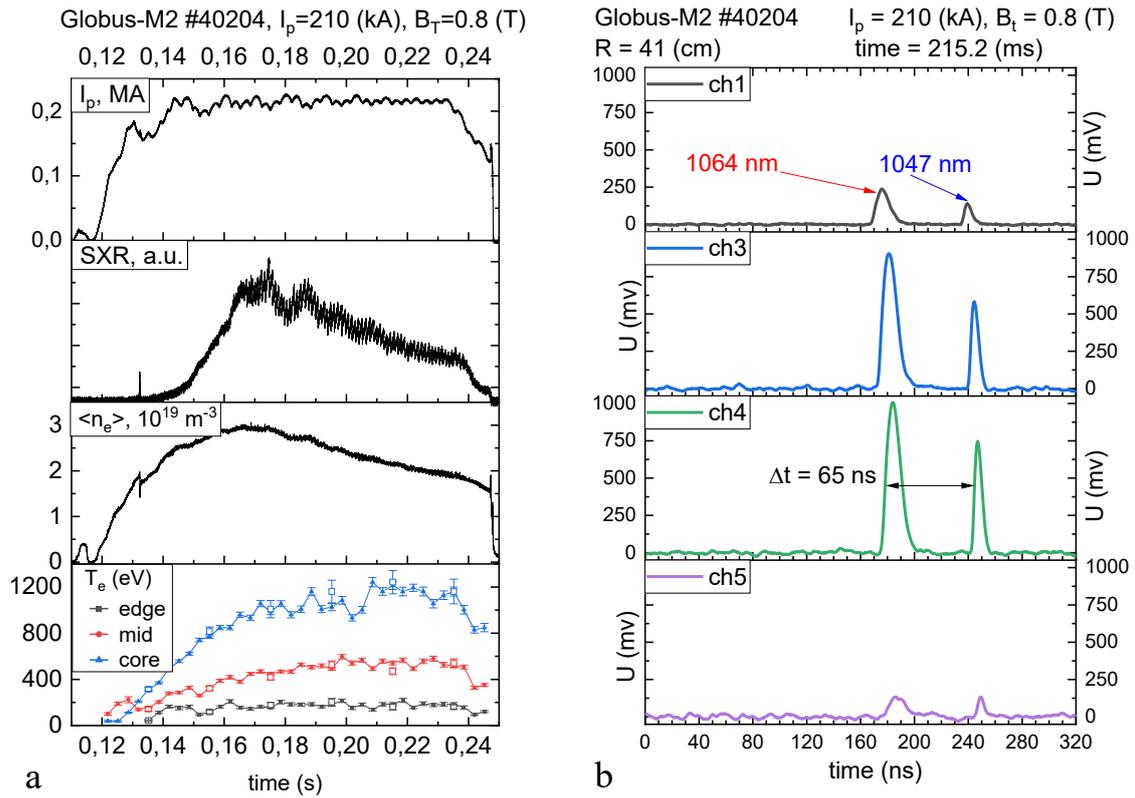

*Figure 17. Multi-colour experiment in the discharge #40204, where multi-colour probing was carried out. a – Time evolution of the main plasma parameters. From top to bottom: plasma current $I_p$, soft X-ray intensity $SXR$, average plasma density $<n_e>_l$, electron temperature $T_e$ for the center (core), edge (edge) of the plasma column and for the middle of a small radius (mid). b – oscillograms of synchronized TS signals for $\lambda_{01}=1064.5$ nm and $\lambda_{02}=1047.3$ nm.*

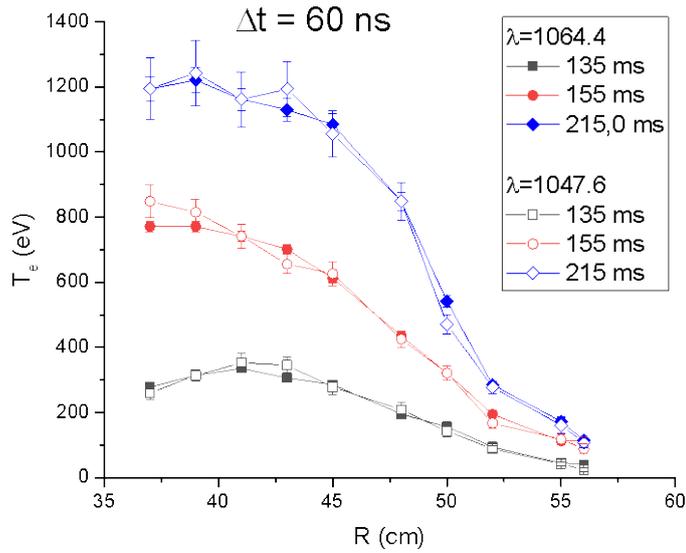

*Figure 18. Electron temperature **T**$_e$ profiles measured in different phases of the tokamak discharge #40204 with a time delay of 65 ns between probing with 1064.5 nm laser (filled markers) and 1047.3 nm laser (empty markers).*

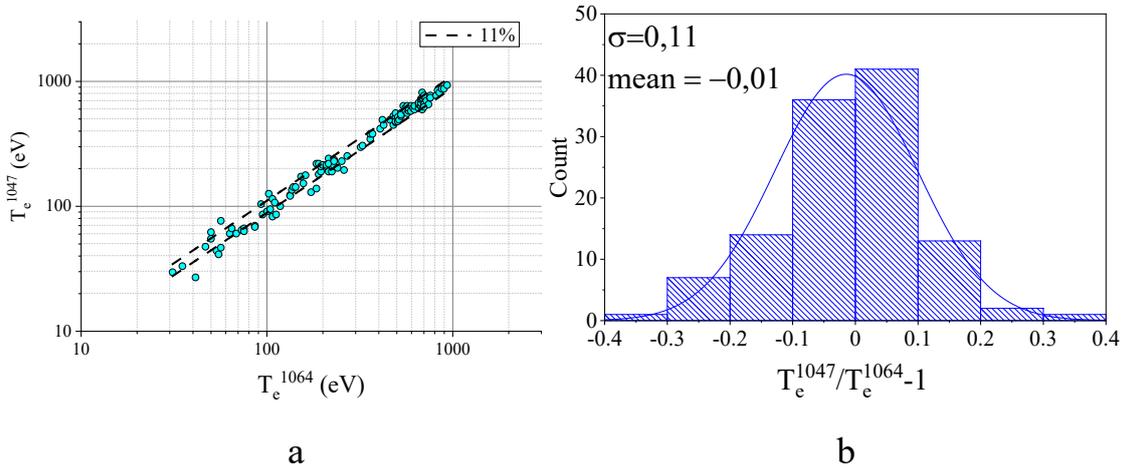

*Figure 19. The multi-colour experiment results, calculated using known spectral calibration of the system. a – Comparison of the electron temperature determined by the classical approach with use of Nd:YAG 1064 nm and Nd:YLF 1047 nm probing lasers, $T_e^{1064}$ on Y-axis and $T_e^{1047}$ on X-axis respectively. b – distribution of the normalized deviation $T_e^{1064}$ from $T_e^{1047}$, the solid line – normal distribution.*

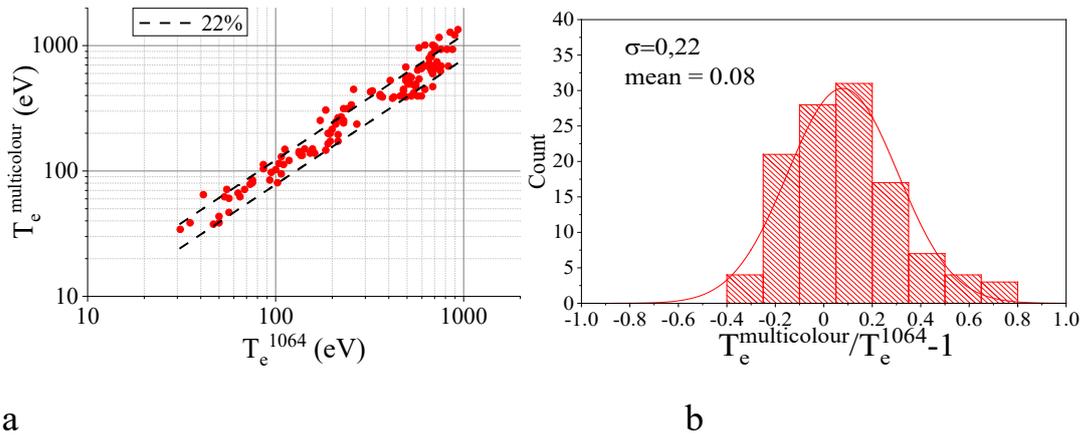

a  b

*Figure 20. The multi-colour experiment results, calculated without using a spectral calibration of the system. a – Comparison of the electron temperature calculated using the multi-colour technique assuming an unknown spectral calibration of the system ($T_e^{mutilcolour}$, Y-axis) with the values determined in the traditional way ($T_e^{1064}$, X-axis). b – distribution of the normalized deviation $T_e^{mutilcolour}$ from $T_e^{1064}$, the solid line – normal distribution.*

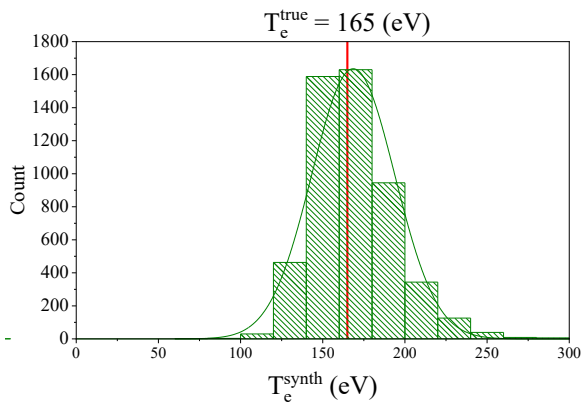 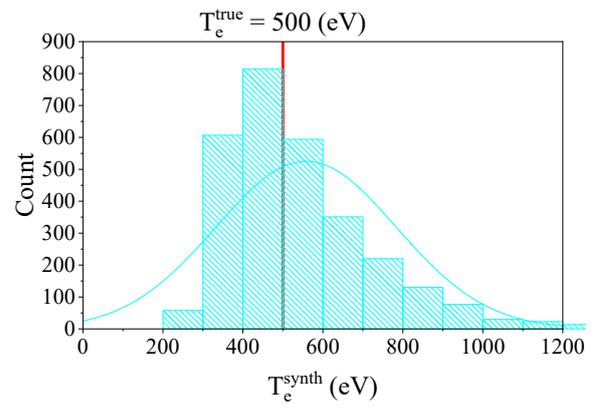

Figure 21. *Distribution of the $T_e^{synth}$ values for given $T_e^{true}$ value of 165 eV (a) and 500 eV (b) calculated in the numerical experiment with unknown spectral calibration of the system. Solid lines – normal distribution.*